\documentstyle[11pt]{article}
\newcommand{\be} {\begin{equation}}

\newcommand{\ee} {\end{equation}}
\newfont{\mc}{cmcsc10 scaled\magstep2}
\newfont{\cmc}{cmcsc10 scaled\magstep1}
\newcommand{\bc}{\begin{center}}
\newcommand{\ec}{\end{center}}
\def\thebibliography#1{\section*{{\bc\normalsize REFERENCES \ec}}\list
 {}{\setlength\labelwidth{1.4em}\leftmargin\labelwidth
 \setlength\parsep{0pt}\setlength\itemsep{0pt}
 \setlength{\itemindent}{-\leftmargin}
 \usecounter{enumi}}
 \def\newblock{\hskip .11em plus .33em minus -.07em}
 \sloppy
 \sfcode`\.=1000\relax}

\begin{document}

\title{{\large{\bf THE DISCOVERY OF 8.7 S PULSATIONS FROM THE\\[-2mm] ULTRASOFT
X--RAY SOURCE 4U~0142+61}}}
\author{\\[-1cm]{\normalsize{\bf {\cmc G. L. Israel$^{\ast}$
}}}\\[-2mm]
\normalsize
International School for Advanced Studies, Via Beirut
2--4,\\[-2mm]
\normalsize
I--34014 Trieste, Italy, {\small israel@vxrmg9.icra.it} \\
{\normalsize{\bf {\cmc S. Mereghetti}}}\\[-2mm]
\normalsize
Istituto di Fisica Cosmica del CNR, Via Bassini 15,\\[-2mm]
\normalsize
 I--20133 Milano,
Italy, {\small sandro@ifctr.mi.cnr.it}\\
{\normalsize{\bf {\cmc and}}}\\
{\normalsize{\bf {\cmc L. Stella\thanks{Affiliated to I.C.R.A.}}}}\\[-2mm]
\normalsize
Osservatorio Astronomico di Brera, Via Brera 28,\\[-2mm]
\normalsize
 I--20121 Milano,
Italy, {\small stella@astmib.mi.astro.it}\\ \\
\date{Accepted for publication on {\it The Astrophysical Journal (Letters)}}}

\maketitle
\pagestyle{empty}
\thispagestyle{empty}
\newpage

\begin{center} \section*{\bf {\normalsize ABSTRACT}}
\vspace{-7mm} \end{center}
\noindent

We discovered  an  X--ray periodicity at $\sim 8.7$~s from the direction of
the two sources 4U~0142+61 and RX J0146.9+6121. The pulsations are visible
only in the 1--4 keV   range, during an observation obtained with   EXOSAT
 in August 1984. In the same data, periodic oscillations at 25 minutes had
been previously found in an additional hard spectral component   above 4
keV. The newly discovered periodicity most likely originates from the
optically unidentified source 4U~0142+61, previously considered a possible
black hole candidate on the basis of its ultrasoft spectrum. Marginal
evidence for the $\sim 8.7$~s pulsations is found in the two 1985 EXOSAT
observations and in a 1991 ROSAT--HRI pointing; if true, these measurements
imply a   spin--up timescale of $\sim 530$~yr. Though the very high
($>$$10^4$) X--ray to optical flux ratio of 4U~0142+61 is compatible with
models based on an isolated neutron star, the simplest explanation involves
a low mass X--ray binary with a   faint companion, similar to 4U~1626--67.
A search for delays in the pulse arrival times caused by orbital motion
gave negative results. The discovery of periodic pulsations from 4U~0142+61
weakens the phenomenological criterion that an ultrasoft spectral component
is a signature of accreting black holes.\\
{\it Subject Headings:} stars: neutron, individual (4U~0142+61) ---
X--rays: sources
\newpage

\vspace{-1cm}
\begin{center} \section*{{\normalsize 1. INTRODUCTION}}\end{center}
\vspace{-3mm}

The persistent X--ray source 4U~0142+61 was discovered by Uhuru and  soon
noticed to possess an ultrasoft spectrum. In the X--ray colour--colour
diagram of White and Marshall (1984) it occupies the same region of   black
hole candidates   in their ``high state", such as LMC~X--3, LMC~X--1 and
A~0620--00. 4U~0142+61 lies in the galactic plane (l=129${\rm^o}$.4,
b=--0${\rm^o}$.4) and, despite its small error circle (a few arcseconds),
no optical or radio counterparts have yet been identified (White et al.
1987). While the X--ray luminosity of 4U~0142+61
($L_x$$\sim$$10^{36}$(d/4kpc)$^{2}$~ergs~s$^{-1}$)
is 1--2 orders of magnitude lower than that
measured for the sources above, its spectrum (power law with energy index
of $\sim$ 3--4) is   reminiscent of high state  black hole candidates.

During one of three EXOSAT observations of 4U~0142+61 carried out in
1984--1985, an additional spectral component was detected above 3~keV within
the $\sim90$~' collimator response of the Medium Energy (ME) experiment.
Correspondingly, $\sim$25~min periodic  oscillations were discovered in the
3--10~keV energy range (White et al. 1987). It is still unclear whether
this component and the  $\sim$25~min oscillations originated from
4U~0142+61 or from a second source in the field of view. Mereghetti, Stella
\& De Nile (1993) pointed out that such a source could be RX~J0146.9+6121,
an  X--ray transient recently discovered with ROSAT and identified with the
Be star LSI~+61${\rm^o}$~235 (Motch et al. 1991).

Here we present the results of a re--analysis of the EXOSAT data, which led
to the discovery of 8.7~s coherent pulsations in the $1-4$~keV band
(Israel, Mereghetti \& Stella 1993).

\vspace{-9mm}
\begin{center} \section*{{\normalsize 2. TIMING ANALYSIS}}\end{center}
\vspace{-3mm}

During the $\sim$$12$~hr observation of  August 27--28, 1984  the average
1--10~keV count rate of 4U~0142+61 in the EXOSAT ME experiment was
$\sim$10.7~counts~s$^{-1}$. Due to the presence of the 3--10~keV spectral
component showing the 25~min oscillations, this rate was $\sim$$40\%$
higher than that measured during the 1985 November 11 and December 11
observations, when the additional component was not present. In the 1984
observation, the ME instrument provided    light curves with a  time
resolution of 1~s in different energy bands. The times were corrected to
the barycenter of the solar system. The power spectrum of the 1--3~keV
light curve over an interval of 32768~s (Fig.~1) revelead the presence of
two highly significant peaks (random occurrence probabilities of
$8.9\times10^{-8}$ and $3.6\times10^{-9}$, respectively), at the
fundamental and the second harmonic  of a coherent modulation with a period
of $8.6872$~s. The power spectrum of the 4--11~keV light curve did not show
any evidence for this pulsation. On the contrary, the peaks corresponding
to the first three harmonics of the ~25~min modulation were clearly visible
in the  4--11~keV power spectrum, but not in the 1--3~keV one  (Fig.~1).

To precisely measure the pulse period, the observation was divided in
intervals of $\sim10^{3}$~s, and for each interval we determined the
relative phase of the 8.7 s pulsations. This was done by fitting with a
Gaussian  the central peak of the cross--correlation function obtained from
the folded light curve of each interval and that of the entire observation.
These phases were then fitted to a linear function giving a best fit period
of $8.68723\pm 0.00004$~s. Introducing a quadratic term did not improve
significantly the fit and allowed to derive a $90\%$ confidence upper limit
to the period derivative of $\dot P < 6.2 \times 10^{-9}$. Between 1 and
4~keV the light curve (Fig.2a) consists of two peaks separated in phase by
 $\sim0.4-0.5$, with a peak to peak amplitude of $\sim15\%$. We note that,
being heavily absorbed (N$_H\sim4\times 10^{23}$~cm$^{-2}$, White et al.
1987), the high energy spectral component does not significantly contribute
below 4 keV.

A search for an orbital modulation of the arrival times of the $8.7$~s
pulses was carried out for 199 orbital periods ranging from 430~s to
43000~s, with a spacing equal to half the Fourier resolution. The data were
folded at the best period for seven different phase intervals of each trial
orbital period. The resulting light curves were cross--correlated with the
average one, and the peak of the cross--correlation function was fitted with
a Gaussian. The centroid of the Gaussian provides an estimate of the delay
of the $8.7$~s pulses in each of the seven phase intervals of a trial
orbital period. A circular orbit would be revealed by a sinusoidal
modulation of the delays. To search for such a modulation, we calculated
the squared Fourier amplitudes of the delays, without finding any
significant deviation from the expected $\chi^2_2$--like distribution. The
$99\%$ confidence upper limit to $a_x \sin$$i$ was derived to be 0.37~lt~s
(this limit reduces to 0.25~lt~s under the  single trial  hypothesis that
the 25~min modulation corresponds to the orbital period). Therefore we
conclude that the EXOSAT data do not provide any evidence for an orbital
motion.

Unfortunately, during the two 1985  observations, energy--resolved ME data
were obtained only with an integration time of 10~s. Only the summed
rates from the ME Argon and Xenon chambers were available with a time
resolution of $0.25$~s. Based on these data, we accumulated 1~s resolved light
curves, which, however  were characterised by a very high level of counting
statistics noise due to the high background from the Xenon chambers
($\sim250$ ct s$^{-1}$). The periodicity was searched in these light curves
using the folding technique. To increase the sensitivity, we considered
only trial periods in the 8.55--8.80~s range, as expected for a rate of
spin change of $| P/\dot P |< 100$~yr (comparable to the highest value
observed from X--ray pulsars) from the 1984 observation. We used   eight
phase bins and a  period spacing which oversampled by a factor of $\sim 20$
the Fourier resolution. Maximum $\chi^2$ of 30.0 and 29.6 were obtained for
the 1985 November 11 and December 11 observations, respectively,
corresponding to periods of $8.6658\pm 0.0005$ and $8.6663\pm 0.0005$~s.
The chance probability of these maxima is difficult to estimate, because
the trial periods are not independent. A lower and an upper limit can be
obtained by considering a number of independent trials equal to the number
of Fourier periods and   the total number of trial periods, respectively.
This gives a probability between $6\times10^{-3}$ and 0.12 for the November
observation and   between $4\times10^{-3}$ and 0.08 for the December one.
For the latter observation the probability reduces to between
$2\times10^{-4}$ and $4\times10^{-3}$ if the range of possible periods is
restricted by assuming that  the November   detection is statistically
significant. In both cases the folded light curves (1--10 keV,  Fig.~2b)
show a double--peaked shape similar to that of the 1984   data, though the
peak to peak amplitudes are larger (30--40$\%$, we note, however, that
these values might be affected by systematic uncertainties in the
background subtraction of the Xenon chambers).

During the three EXOSAT ME observations, simultaneous imaging in the
0.05--2~keV band was obtained with the  low energy  (LE) telescope. For each
observation, we folded the LE light curve  of 4U~0142+61 at the
corresponding  period determined from the ME data, without finding any
evidence of modulation. However, due to the small counting statistics in
the LE instrument, the derived upper limits on the peak to peak amplitudes
are not very constraining ($TBD\%$, 34$\%$ and 44$\%$ respectively for the
1984, November 1985 and December 1985 observations, 99$\%$ confidence
level).

A  2180~s long ROSAT observation of 4U~0142+61, was carried out on February
13, 1991 with the HRI instrument (0.1--2.4 keV). After correction to the
solar system barycenter, the arrival times of the 2846~counts within a
radius of 20"   from the   position  of 4U~0142+61 were searched for
periodicities using the Rayleigh test (Leahy, Elsner and Weisskopf 1983).
We considered 87 independent periods between 8.0 and 9.4~s, finding a
maximum value of the  Rayleigh test statistics of 17.23  for
P=$8.600\pm0.017$~s (chance probability   $\sim1.6\%$). The folded light
curve shows a broad, almost sinusoidal modulation, with a peak to peak
amplitude of $\sim 30\%$ (see Fig.2b).

In view of the relatively high probabilities of chance occurrence, the
detection of the periodicity in the 1985 and 1991 data cannot be considered
certain. It is, nevertheless, intriguing that the four period values are
consistent with a line of constant spin--up, on a timescale of
$\simeq530$~yr.

\vspace{-5mm}
\begin{center}\section*{\normalsize 3. DISCUSSION}\end{center}
\vspace{-3mm}

Since the 8.7 s  periodicity has been detected at a high confidence level
only in   non--imaging data, we cannot   exclude that it originates in a
source different from  4U~0142+61. The only other  X--ray source presently
known in this region is the Be star LSI~+61${\rm^o}$~235, probably a binary
system containing an accreting compact object (Motch et al. 1991;
Mereghetti et al. 1993). If the latter spins at 8.7~s, a reasonable value
for a neutron star accreting from a Be companion, the 25 min periodicity
remains to be explained. It could result from an orbital modulation in a
low mass X--ray binary (4U~0142+61), or from (quasi)--periodic flares in
LSI~+61${\rm^o}$~235, similar to those observed in other Be systems (Parmar
et al. 1989, Finley et al. 1992), but both possibilities present with some
difficulties (see  White et al. 1987 and Mereghetti et al. 1993). Though a
third, yet unknown, source in the field of view could help in solving the
puzzle,  we regard this as an unlikely {\it ad hoc} explanation, and
therefore assume in the following that the 8.7 s pulsations are due to
4U~0142+61. The lack of an optical counterpart down to limits of
V~$\sim$24 and R~$\sim$ 22.5 (Steinle et al. 1987, White et al. 1987)
implies an X--ray to optical flux ratio,  $F_x/F_{opt}$$>$$10^4$. The only
known classes of galactic sources which can yield such a high $F_x/F_{opt}$
value are low mass X--ray binaries (LMXRBs) and isolated neutron stars.

\begin{center}
\subsection*{{\normalsize 3.1. {\it A low mass X--ray binary ?}}}
\end{center}
\vspace{-3mm}

A neutron star accreting from a low mass companion is the most likely
explanation for 4U 0142+61, especially if the spin--up evidence is
confirmed by further observations. Coherent  pulsations are rarely seen in
LMXRBs: the only known examples among optically identified systems,
4U~1626--67, Her~X--1 and GX~1+4, have very different  X--ray properties,
companion stars and evolutionary origins (see, e.g., White, Nagase \&
Parmar 1993). The spin period of 4U~0142+61 is very similar to that of
4U~1626--67 (7.7~s, Rappaport et al. 1977), and it is interesting to note
that two other optically unidentified pulsars,  which are likely
accreting from low mass companions, 1E~2259+586  and 1E~1048.1--5937
(Coe \& Jones 1992; Mereghetti, Caraveo \& Bignami 1992), have periods
of the same order, 6.98 and 6.44~s respectively (Davies et al. 1990;
Corbet \& Day 1990).

The position of 4U~0142+61 is close ($<$0.5${\rm^o}$) to that of two open
clusters with well determined distances and reddening: NGC 654 at 2.5 kpc
(A$_V$=2.67), and NGC 663 at 2.1 kpc (A$_V$=2.43)  (see Leisawitz, Bash and
Thaddeus 1989, and references therein). The column density
N$_H\sim1.5\times 10^{22}$~cm$^{-2}$ derived from the power law spectral
fits of 4U~0142+61 (White et al. 1987) corresponds to a higher absorption,
A$_V$$\sim$7  (Gorenstein 1975), hinting to a greater distance. However,
4U~0142+61 is not necessarily   much further than these clusters, since a
part of its absorption could be intrinsic or due to a local (d$<$1~kpc)
molecular cloud which is present in this region, as clearly visible on the
POSS prints. 4U~0142+61 lies near to the edge of this cloud, which does not
significantly affect NGC 654 and NGC 663 (Leisawitz et al. 1989). A
distance of 4~kpc would yield to a 1--10~keV luminosity of $\sim
10^{36}$~ergs~s$^{-1}$, similar to  4U~1626--67. At this distance and
reddening the   faint optical counterpart of the latter source would be
fainter than the present limits for 4U~0142+61. On the other hand, an
evolved companion similar to that of GX~1+4 or Cyg~X--2 (M$_V$$\sim$~--1,
van Paradijs 1991) would have been detected even at $\sim$10~kpc (which for
this direction is well outside the Galaxy). A companion star similar to
that of 4U~1626--67, i.e. either a main sequence star with M$\simeq 0.08
M_{\odot}$ or a white dwarf of 0.02 $M_{\odot}$ (Verbunt, Wijers \& Burm
1990), is also compatible with the limits on $a_x \sin i$ derived in the
previous section, which however also allow for more massive companions. For
instance, a hydrogen main sequence star of $\sim$0.3~$M_{\odot}$, would
fill the Roche lobe for an orbital period of $\sim3$ hours, requiring {\it
i} $<46{\rm^o}$.

Despite the above similarities with  4U~1626--67, there are also important
differences. First, the X--ray spectrum  is much softer than that of
4U~1626--67, a  flat power law (energy index $\sim$0.4) with a cut--off at
$\sim$$20$~keV, similar to that of most accreting X--ray pulsars (White,
Swank \& Holt 1983). In this respect 4U~0142+61 is more similar to
1E~2259+586, whose spectrum can be described by a power law with energy
index $\sim$3, plus some possible cyclotron features suggesting a magnetic
field B $\sim5\times$$10^{11} G$  (Iwasawa, Koyama \& Halpern 1992).
Second, the flux in the ultrasoft spectrum of 4U~0142+61 is rather stable,
unlike most accreting X--ray pulsars, and 4U~1626--67 in particular  which
shows quasi--periodic flares on a timescale of 1000~s (Li et al. 1980).
Finally, the spin--up timescale of $\sim$530~yr, if confirmed, would be
about a factor ten shorter than that observed in  4U~1626--67. In the
standard framework of accretion disk torques on magnetized neutron stars
(see, e.g., Henrichs 1983) this would require a mass accretion rate of
$\sim 3 \times 10^{17}(B/10^{12}$ G$)^{-1/3}$~g~s$^{-1}$. Such a  high
accretion rate is not incompatible with the flux measured from 4U~0142+61,
since the accretion luminosity can easily be two orders of magnitude higher
than that observed in the 1--10 keV range, if the steep spectrum   extends
down to $\sim 0.1$~keV.

\vspace{-7mm}
\begin{center}
\subsection*{{\normalsize 3.2. {\it An isolated neutron star ?}}}\end{center}
\vspace{-3mm}

The   possibility that  4U~0142+61 is an isolated neutron star is suggested
by its very high  $F_x/F_{opt}$, its ultrasoft spectrum, and the absence of
significant variability on long timescales (of course this possibility
requires that the evidence for secular spin--up is the result of chance
detections in the 1985 and 1991 data). In principle the X--ray emission
could be due to non--thermal magnetospheric processes powered by the
rotational energy, to thermal emission from the neutron star surface, or to
accretion from the interstellar medium. While examples of the first two
mechanisms are well known (see, e.g., Mereghetti, Caraveo \& Bignami 1994),
no compelling evidence for a compact object accreting from the interstellar
medium has yet been found, despite several studies show that such sources
could be relatively common (Treves \& Colpi 1991; Blaes \& Madau 1993).

For a spin period of 8.7 s and any reasonable magnetic field
($\leq10^{14}$~G), the available rotational energy loss is too small, unless
4U~0142+61 is at a distance of  a few parsecs. To investigate the
possibility of thermal emission, we fitted blackbody spectra to the 1985
ME+LE data, obtaining  kT$\sim 0.5$~keV,
N$_H\sim$2--4$\times10^{21}$~cm$^{-2}$ and a bolometric luminosity
$\sim10^{32}~(d/100~pc)^{2}$~ergs~s$^{-1}$ (however these fits were
substantially worse than those with a power law:  reduced $\chi^2$ of
$\sim$3--6). This implies an emission region of
$\sim0.1~(d/100\ {\rm pc})^{2}~~{\rm km}^{2}~$,
compatible with a hot spot on the surface of a
neutron star, possibly the magnetic polar cap heated by accretion  from the
interstellar medium. In this case the accretion induced luminosity would be
$\sim10^{32}\ (v/50~km~s^{-1})^{-3}\ (n/100~cm^{-3})\ \ {\rm ergs~s}^{-1}$,
where  {\it v} is the neutron star velocity relative to the interstellar
medium of density {\it n} (Blaes \& Madau 1993). The already mentioned
molecular cloud (Leisawitz et al. 1989) could be at a distance of only a
few hundred parsecs and easily provide the density required to power the
observed luminosity. For accretion onto the neutron star to take place, the
magnetospheric centrifugal barrier must be open, and the low rates implied
by the above luminosity therefore require a magnetic field  $B$$ \leq$
$10^{10}~(d/100\ {\rm~pc})$~G (see, e.g., Stella et al. 1994).

\vspace{-1cm}
\begin{center}\section*{\normalsize 4. CONCLUSIONS}\end{center}
\vspace{-3mm}

The discovery of 8.7~s pulsations further complicates the puzzle of the
X--ray emission from the region of sky containing 4U~0142+61 and
LSI~+61${\rm^o}$~235/RX~J0146.9+6121. The most likely origin for the newly
discovered periodicity is the ultra soft source 4U~0142+61, which is
probably a LMXRB with a very faint companion, similar  to  4U~1626--67. In
the absence of an optical identification for  4U~0142+61, this would be
supported if the evidence for the secular spin--up reported  in this {\it
Letter} were confirmed. Independent  of the   nature of  4U~0142+61, we
note that the detection of coherent pulsations  weakens the
phenomenological criterion that the presence of an ultrasoft X--ray spectral
component is a characteristic of accreting black holes (White \& Marshall
1984).

\vspace{5mm}

Acknowledgements.

The data were obtained through the High Energy Astrophysics Database
Service at the Astronomical Observatory of Brera and the EXOSAT interactive
analysis system at ESTEC. We are grateful to A.Reynolds and A.Parmar for
help with the data processing. This work has been partially supported by
ASI grants.

\newpage

\vspace{6cm}
{\bf Figure 1}: Power spectra of   the 1984
EXOSAT observation in two different energy ranges.
The peaks corresponding to the first two
harmonics of the $\sim 8.7$~s pulsations are clearly visible in the 1--3~keV
power spectrum (upper panel). The three peaks in the 4--11~keV
power spectrum testify to
the presence of the $\sim$25--min modulation in that energy range.
\vspace{4mm}

{\bf Figure 2}: {\it (a)} Folded light curves of the $\sim 8.7$~s pulsations in
four different energy bands during the 1984 EXOSAT observation.
{\it (b)}  Light curves folded at the most significant period of the
two 1985 EXOSAT observations and the 1991 ROSAT observation.

\end{document}